# The sensitivity of the microstructure and properties to the peak temperature in an ultrafast heat treated low carbon-steel

M.A. Valdes-Tabernero[1*], A. Kumar[2], R.H. Petrov[2,3], M.A. Monclus[1], J.M. Molina-Aldareguia[1], I. Sabirov[1]

[1] IMDEA Materials Institute, Calle Eric Kandel 2, Getafe 28906, Madrid, Spain

[2] Department of Materials Science and Engineering, Delft University of Technology, Mekelweg 2, 2628CD Delft, The Netherlands

[3] Department of Electrical Energy, Metals, Mechanical constructions & Systems, Ghent University, Technologiepark 46, 9052 Ghent, Belgium

*Abstract*

In this work, we investigate the sensitivity of the microstructure and mechanical properties of an ultrafast heat treated low carbon-steel to the peak temperature. In all studied cases, the steel was heated within the intercritical temperature range (i.e. between the $A_{C1}$ and $A_{C3}$ temperatures). Both the peak temperature and soaking time were varied, and their effect on the size, the fraction of individual microstructural constituents and their tensile mechanical response were investigated. It is shown that the increasing peak temperature and soaking time promote austenite formation and recrystallization processes in the ferritic matrix. The highest nanohardness is shown by martensitic grains, while recovered ferrite demonstrated slightly higher nanohardness compared to recrystallized ferrite. The applied heat treatment parameters have strong effect on the nanohardness of martensite, whereas nanohardness of ferrite microconstituents is not sensitive to variation of the peak temperature and soaking time. The non-recrystallized ferrite is harder than its recrystallized counterpart due to the higher dislocation density of the former. Increasing

---

[*]Corresponding author: Miguel Angel Valdés Tabernero.

Postal address: IMDEA Materials Institute, Calle Eric Kandel 2, Getafe 28906, Madrid, Spain.

Phone: +34 91 5493422. E-mail: miguelvaldestabernero@gmail.com

peak temperatures promote strengthening in the material at the expense of its ductility mainly due to increased martensite fraction. The steel demonstrates enhanced strain hardening ability independently of the peak temperature. Analysis of the experimental results showed that the industrial processing window of ± 10 °C may lead to some heterogeneity of the local microstructure in the ultrafast heat treated sheets. However, the latter should not have any negative effect on the overall mechanical behavior of the ultrafast heat treated steel sheets on the macro-scale.

1. **Introduction**

Steel sheets manufacturing is a multistage process, where the steel is subjected to several rolling operations and finally to a heat treatment, which determines its final microstructure and, therefore, its properties. The standard approach for processing advanced high strength steels (AHSS) is based on the homogenization of the microstructure at elevated temperatures followed by cooling with well controlled rates [1]. The typical route used to manufacture components for the automotive industry, where steel is the primarily used material [2], is a relatively long process resulting in very high energy consumption and carbon dioxide emission [3]. A potential solution to decrease the $CO_2$ emissions produced by vehicles is to reduce the total weight of the car, without compromising the passengers' safety. In order to do so, the mechanical properties of the car components should be improved. Therefore, the steel industry is continuously looking for new solutions to fulfill the current societal demands by processing the steel in the most environmentally-friendly manner. Hence, in the last decades, new processing routes were developed, such as the rapid or "flash processing" treatment [4,5]. In literature, this process is also referred to as "ultra-fast heating" (UFH) [6,7] or "ultra-rapid annealing" (URA) [8,9]. It is based on heating the material to intercritical or fully austenitic temperature with heating rates well above 100 ºC/s, which is at least one order of magnitude higher than the conventional heating rates (≤ 10 ºC/s), followed by a short soaking at peak temperature and immediate quenching to the room temperature. Thus, the treatment time and the energy consumed for the process are significantly reduced [10].

Complex multiphase microstructure consisting of ferrite and martensite and some retained austenite is typically formed after UFH treatment. The resulting microstructure and properties of the UFH-treated material are greatly affected by the initial microstructure as well as the heat treatment parameters, such as heating rate, soaking time and peak temperature [11]. It has been reported that high heating rates increase the austenite start and finish temperatures [12,13]. Moreover, in low carbon steels, recrystallization temperature tends to increase with increasing heating rate and may even exceed the austenite start temperature ($A_{C1}$) [14,15]. Hence, ferrite to austenite transformation takes place in a non-recrystallized matrix, which leads to the formation of numerous austenite nuclei, while the ferritic matrix undergoes recrystallization and recovery simultaneously. Thus, the complex microstructure with finer grain size is developed [16]. In order to promote grain refinement, the isothermal soaking time is typically kept as short as possible – in the 0.1-0.2 s range [7,11,15,17]. However, such soaking times constitute a challenge for the steel sector due to its difficult implementation in the current industrial lines, impeding the expansion of this processing route. Slightly longer soaking times (1 – 3 s) can be a feasible solution to maintain the reduced grain size brought about by ultrafast heating [18], but only if the effect of other processing parameters is well understood. One key parameter is the peak temperature, as it influences the mechanism of the austenite formation and growth. At conventional heating rates the austenite formation kinetics are determined by carbon diffusion, whereas at ultrafast heating rates formation of austenite starts by carbon diffusion control, which is later overtaken by a massive mechanism [7,9,19]. The martensite formed after quenching gets softer due to higher volume fraction of intercritical austenite and reduced carbon content therein [20]. Additionally, the recrystallized ferrite volume fraction in the heat treated steel tends to increase with increasing peak temperature [21]. Despite significant effect of peak temperature on the final microstructure of the UFH treated steels, there are no systematic studies focused on the microstructure sensitivity to the peak temperature variations. In addition, it is important to simulate real industrial conditions, as the typical industrial processing temperature window of ± 10 ºC is by an order of magnitude higher compared to that maintained in a laboratory dilatometry or thermo-mechanical simulator (± 1 ºC) [7,11]. Therefore, it is essential to understand the effect of peak temperature ranges during UFH on the microstructure and properties in order to select the optimum conditions.

The reported studies on the UFH treatment of steels have mainly focused on microstructure and basic mechanical properties (hardness, tensile strength, ductility). In references [17,22], it was reported that the UFH treatment leads to an improvement on the material strength compared to the conventional heat treatment, without a reduction in ductility. However, although it has been shown in many studies that mechanical behavior of multi-phase materials on macroscale depends on the morphology, architecture and properties of the individual microconstituents [1,23,24], there are no such in-depth studies on the UFH treated steels. Hence, understanding the *heat treatment parameters-microstructure-properties* relationship both at macro- and micro-scales, is necessary to develop specific microstructures and properties and to design and optimize precise UFH treatments depending on the requirements and specifications of the final product. Therefore, the objective of this work is to investigate the influence of peak temperature and short soaking times (≤ 1.5 s) on the microstructure and properties of the individual microstructural constituents, as well as to relate both to the macro-mechanical response of the material.

## 2. Material and experimental procedures

### 2.1. Material

The chemical composition of the low carbon steel selected for this investigation was 0.19 % C, 1.61 % Mn, 1.06 % Al, 0.5 % Si (in wt. %). The as-received material was 1 mm thick sheet (50% cold rolled) with a microstructure of 76 % of ferrite and 24 % of pearlite (**Figure 1**). This material was subjected to two kinds of heating experiments: a) dilatometry measurements to determine the formation of austenite at different intercritical temperatures, and b) annealing tests to different intercritical temperatures with varying soaking time followed by quenching. Both types of experiments are described in detail below.

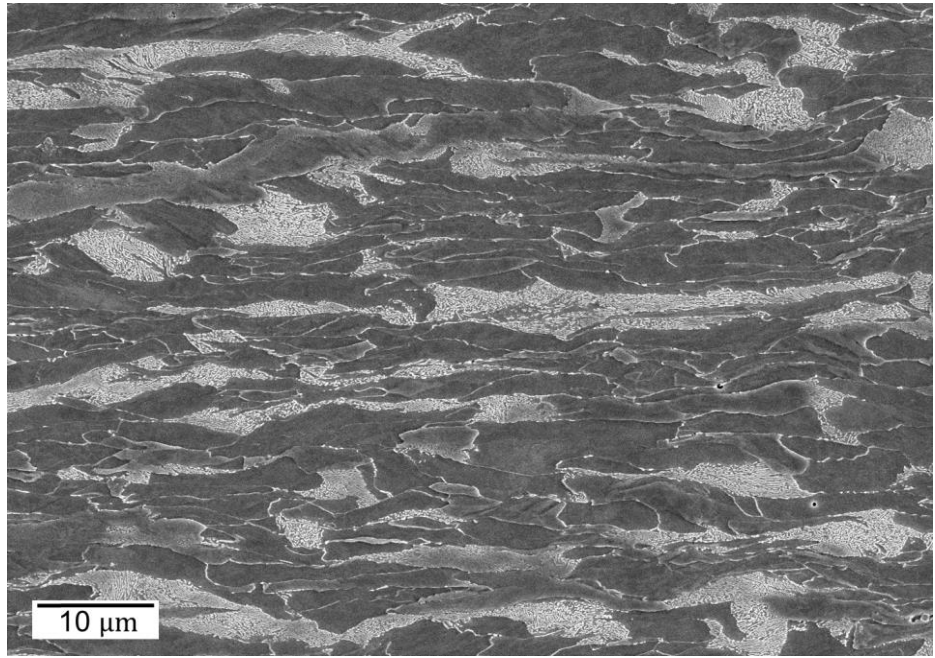

**Figure 1**: Initial ferritic-pearlitic microstructure of the material after 50 % cold reduction.

*2.2. Dilatometry experiments*

Dilatometry measurements were carried out to analyze the austenitization kinetics at different temperatures for the same heating rate. For these experiments, specimens with dimensions of 10x5x1 mm$^3$ were machined from the as-received material. Tests were carried out in a Bähr DIL805A/D dilatometer (Bähr-Thermoanalyse GmbH, Hüll-Horst, Germany). A *K*-type thermocouple was welded to the midsection of each specimen to control their temperature during the experiment. Specimens were heated from room temperature to different temperatures in the intercritical region (860 ºC, 880 ºC and 900 ºC) at 200 ºC/s and soaked for 600 s. Then, specimens were heated to a maximum temperature of 1100 ºC at 200 ºC/s and soaked for 0.2 s (to ensure full austenitization) followed by quenching to room temperature at -300 ºC/s. The volume fraction of the austenite phase formed during isothermal holding was obtained via analysis of the dilatometry data applying the lever rule to the dilatation-time curve [25].

*2.3. Intercritical heat treatment*

Strips having a length of 100 mm and width of 10 mm were cut along the rolling direction from the cold rolled sheet. A *K*-type thermocouple was spot-welded to the midsection of each strip. A thermo-mechanical simulator Gleeble 3800 was used to perform heat treatments. At the first stage of heat treatments, samples were heated at 10 ºC/s to 300 ºC and kept at this temperature for 30 s to simulate a preheating in some industrial continuous annealing lines to minimize the thermal stresses during heating. At the second stage, part of samples was heated from 300 ºC at 800 ºC/s (which corresponds to the ultrafast heating rate) to the intercritical peak temperature of 860 ºC followed by soaking for 0.2 s or 1.5 s and quenching to room temperature with cooling rate of ~160 ºC/s. Hereafter, these specimens will be referred to as UFH860-0.2s and UFH860-1.5s, respectively.

To investigate the influence of peak temperature, additional heat treatments with maximum temperatures equal to 880 ºC and 900 ºC and same soaking times (0.2 s and 1.5 s) were performed. These conditions are referred to as UFH880-0.2s and UFH880-1.5s for the 880 ºC, and UFH900-0.2s and UFH900-1.5s for the 900 ºC heat treatment. Microstructural analysis and hardness measurements along the axis of the heat treated strips showed a homogeneously heat treated zone having a length of 10 mm. The specimens processed by the Gleeble thermo-mechanical simulator were then subjected to a thorough microstructural and mechanical characterization.

## 2.4. Microstructural characterization

Scanning electron microscopy (SEM) and electron backscatter diffraction (EBSD) analyses were carried out for a thorough microstructural characterization of the heat treated samples. Specimens were ground and polished to a mirror-like surface applying standard metallographic techniques with final polishing using OP-U (colloidal silica). For SEM characterization, polished specimens were etched with 2 vol.% nital solution for ~10 s. The EBSD studies were performed using a FEI Quanta™ Helios NanoLab 600i equipped with a NordlysNano detector controlled by the AZtec Oxford Instruments Nanoanalysis (version 2.4) software. The data were acquired at an accelerating voltage of 18 kV, a working distance of 8 mm, a tilt angle of 70º, and a step size of 65 nm in a hexagonal scan grid. The orientation data were post-processed using HKL Post-processing Oxford Instruments Nanotechnology (version 5.1©) software and TSL Data analysis version 7.3 software. Grains were defined as a minimum of 4 pixels with a misorientation $\geq$ 5º. Grain boundaries having a misorientation $\geq$ 15º were defined as high-angle grain boundaries (HAGBs), whereas low-angle grain boundaries (LAGBs) had a misorientation < 15º. The volume fractions of transformed/untransformed grains and recrystallized/recovered ferritic grains were determined by a two-step partitioning procedure described in [17,26]. In this procedure, grains with high (> 70º) and low (≤ 70º) grain average image qualities are separated in a first step, allowing to distinguish between untransformed (ferrite) and transformed (martensite) fractions, respectively. In the second step, recrystallized and non-recrystallized ferritic grains are separated using the grain orientation spread (GOS) criterion: Grains with GOS below 1º are defined as the recrystallized grains, while grains with GOS above 1º are defined as the non-recrystallized ones [27]. Microstructure was observed on the plane perpendicular to the sample transverse direction (the RD–ND plane).

*2.5. Mechanical characterization*

HysitronTI950 Triboindenter with a Berkovich tip was employed for nanoindentation testing. First, square areas having a size of ~10 x 10 μm$^2$ were analyzed by EBSD, and individual microstructural constituents were determined. At least ten areas were tested for each material's condition. In order to target specific phases/grains, these square areas were scanned, using the scanning probe microscopy (SPM) mode of the instrument prior to the nanoindentation. In SPM mode, the nanoindenter tip is in contact with the surface of the tested material scanning it, giving the topography of the sample. Nanoindentation tests were carried out in displacement control mode at a constant strain rate ($\dot{\varepsilon}=\dot{h}/h$) of 0.07 s$^{-1}$, where h is the penetration depth and $\dot{h}$ the penetration rate of the indenter. At least 20 indents were performed on each phase at an imposed maximum depth of 150 nm. The nanohardness was determined from the analysis of the load–displacement curves using the Oliver and Pharr method [28].

Vickers hardness tests of all heat treated samples were carried out using Shimadzu HMV hardness tester according to the ASTM E92 – 17 Standard. The RD–ND plane of samples was ground and polished using 1 μm diamond paste at the final stage. A load of 4.9 N was applied for 15 s.

A Kammrath&Weiss module was used for tensile testing of dog bone sub-size samples at room temperature at a constant cross head speed corresponding to an initial strain rate of 10$^{-3}$ s$^{-1}$. These samples had a gauge length of 4 mm, a gauge width of 1 mm and a thickness of 0.9 mm. They were machined from the homogeneously heat treated zone of the heat treated strips, so their tensile axis was parallel to the RD. All samples were carefully ground and mechanically polished using OP-U (colloidal silica) at the final stage. At least three specimens were tested for each condition, and the results were found to be reproducible.

## 3. Results

*3.1. Microstructural characterization*

*3.1.1. Dilatometry*

**Figure 2** represents the evolution of austenite fraction during isothermal holding at different intercritical temperatures. It is seen that the higher the peak temperature, the higher the initial austenite fraction, as the material is closer to the $A_{C3}$ temperature. For instance, at 860 ºC the austenite volume fraction is 19 %, and increases to 47 % with increasing peak temperature to 900 ºC. It is also seen that the peak temperature strongly affects the kinetics of austenite formation and growth. Austenite rapidly grows at the early stages of annealing at 900 ºC compared to annealing at lower peak temperatures of 860 ºC and 880 ºC, which present a similar behavior. Soaking at 900 ºC for 600 s is sufficient for full austenitization, which is reached after 522 s. At 880 ºC, the austenite fraction achieved at the end of the intercritical holding is 99 %, whereas at 860 ºC is 94 %. Nevertheless, taking into account the positive slope of the curve, it is clear that the material will reach the complete austenitization after soaking for longer time.

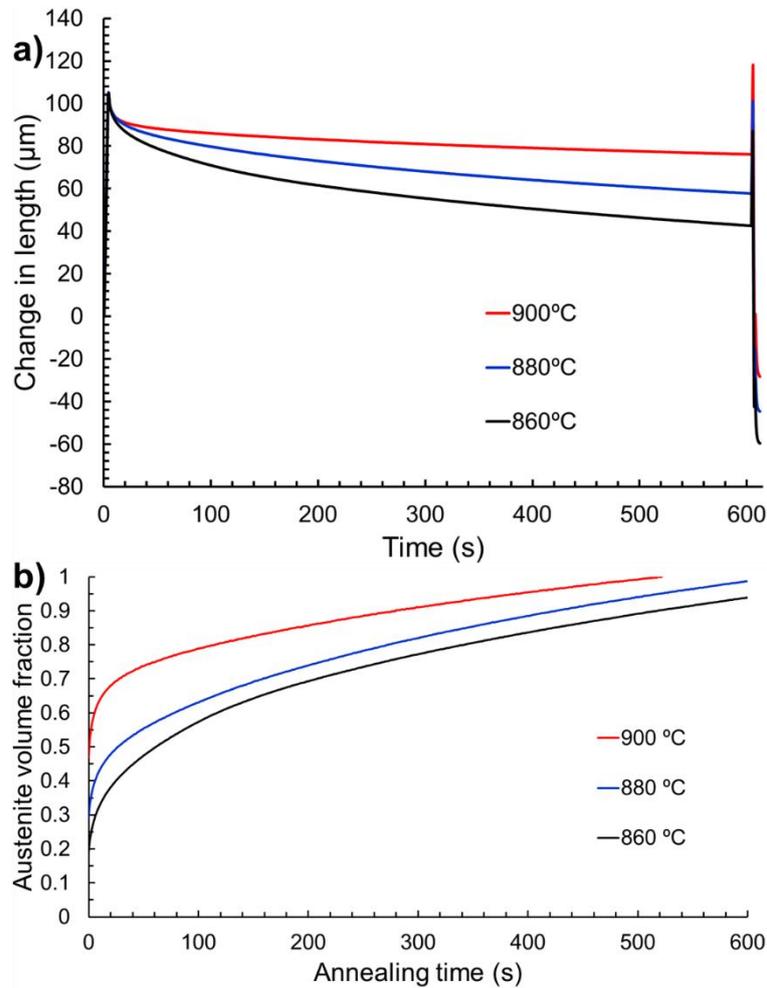

**Figure 2**: a) Dilatation-time curves for material heated to 1100 ºC at 200 ºC/s with soaking for 600 s at different temperatures (860 ºC, 880 ºC and 900 ºC); b) Effect of the peak temperature on the austenite volume fraction during isothermal holding.

### 3.1.2. SEM analysis

SEM analysis of the Gleeble processed samples was performed to qualitatively characterize the influence of both peak temperature and short soaking times on the microstructure. **Figure 3** displays the microstructure variation at the different peak temperatures studied for holding times of 0.2 s (**Figure 3** a-c) and 1.5 s (**Figure 3** d-f). The resultant microstructure is heterogeneous after all heat treatments being mainly formed by ferritic matrix and martensite (marked by a white dashed arrow in **Figure 3** b). In turn, the matrix is formed by recrystallized (Rx) and non-recrystallized (non-Rx) ferrite, as it is demonstrated in Section 3.1.3. A qualitative analysis shows that, independently on the heat treatment parameters, all the conditions present a ferritic matrix consisting of coarse and fine grains due to combination of different processes, which take place during UFH (recovery, incomplete recrystallization and grain growth at early stages). Images on **Figure 3** a-f demonstrate that the increasing peak temperature leads to grain growth for both microstructural constituents (ferrite and martensite) even after a holding time of 0.2 s. Moreover, it is possible to observe that the martensite fraction substantially grows with increasing soaking time independently of the peak temperature and also with the peak temperature for a specific soaking time. As described in Section 3.1.1, increasing peak temperature and soaking time lead to a higher fraction of intercritical austenite, which is transformed into martensite upon quenching. The higher the intercritical austenite fraction, the lower its carbon content due to the C redistribution in its interior. A small amount of retained austenite grains was also identified by EBSD analysis in all conditions (see Section 3.1.3). In addition, very small amount of spheroidized cementite was observed in the microstructure, which remains in the material from the initial cold rolled state (marked by red arrows on **Figure 3** d). Its presence is related to the short time given for its dissolution, hence, it is more commonly observed in the samples annealed for the shorter soaking time of 0.2 s.

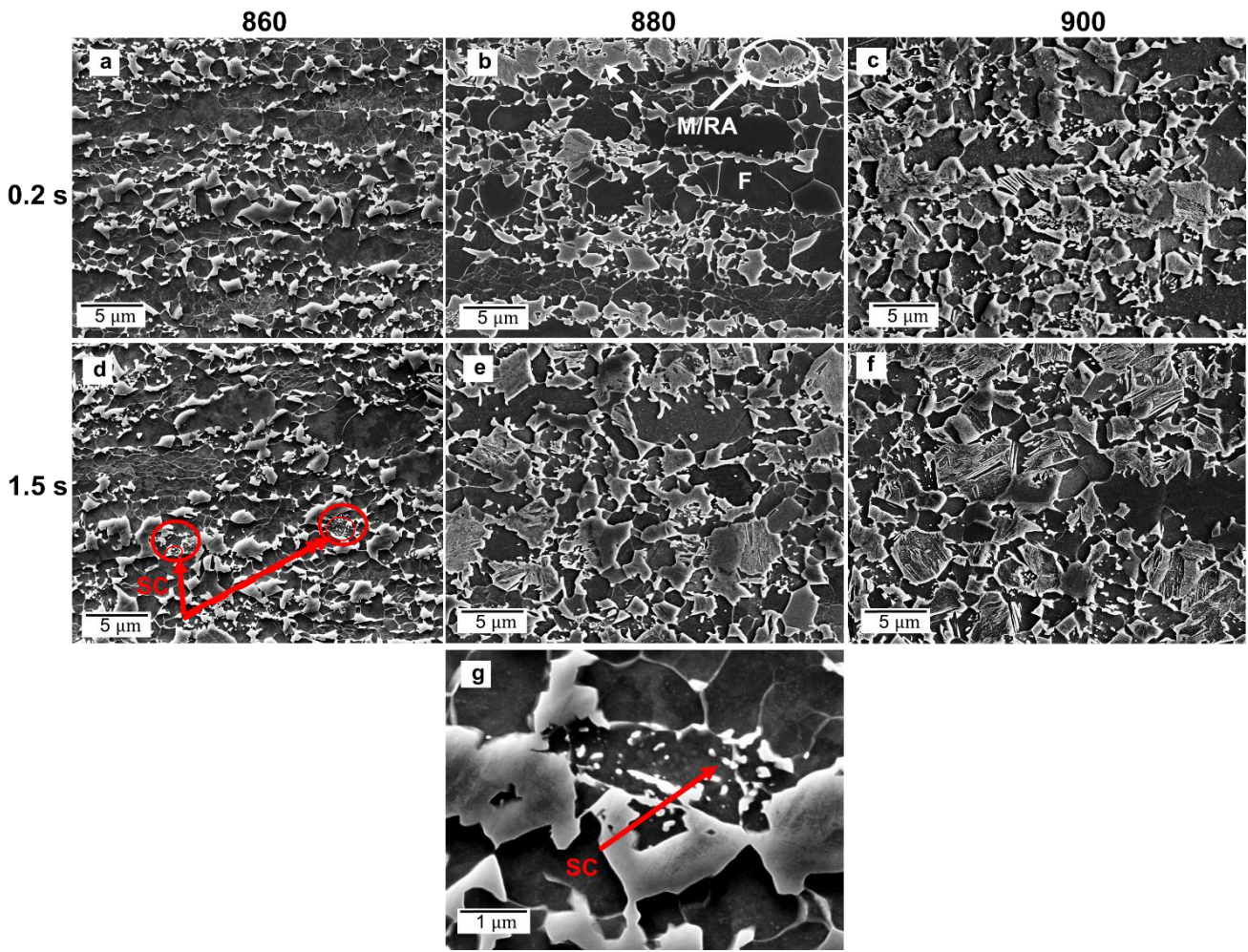

**Figure 3**: SEM photos illustrating the influence of peak temperature (860, 880 and 900 °C) and soaking time (0.2 and 1.5 s) on the microstructure: a), b) and c) are for 0.2 s at 860, 880 and 900 °C, respectively; d) & g), e) and f) are for 1.5 s at 860, 880 and 900 °C, respectively.
SC: spheroidized cementite; M: martensite; F: ferrite; RA: retained austenite.

### 3.1.3. EBSD analysis

EBSD analysis was performed in order to identify and quantitatively characterize the different phases present in the microstructure of the heat treated samples. **Figure 4**a illustrates a typical EBSD phase map measured on the UFH860-0.2s sample. Fine retained austenite grains (in white color) and martensite grains (in black color) are embedded into the ferrite matrix composed of recrystallized (Rx) ferrite (in orange color) and non-recrystallized (non-Rx) ferrite (in blue color). LAGBs are seen mainly in the interior of the non-Rx ferrite grains, whereas majority of the Rx ferrite grains are free of LAGBs. The morphology of the microstructure and the individual microconstituents are very similar in all studied conditions, whereas size and volume fraction of individual microconstituents depend on the heat treatment parameters. The results of the quantitative analysis are summarized in **Table 1**.

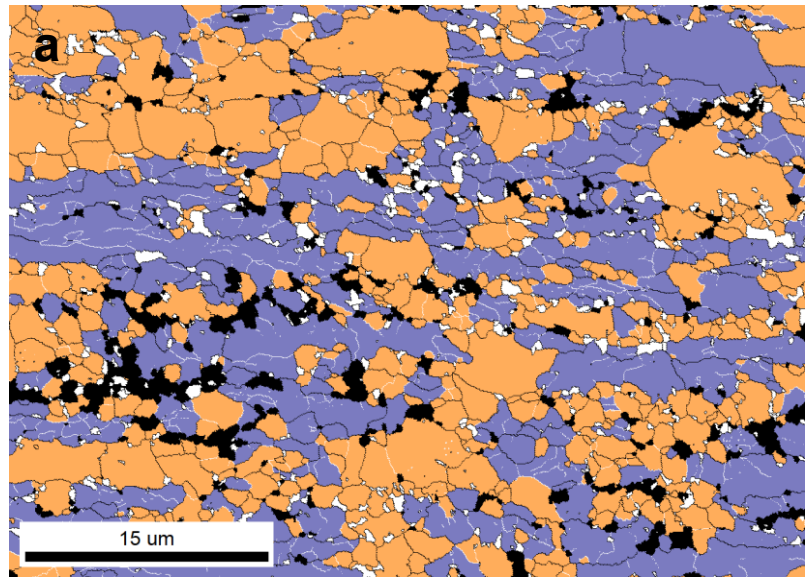

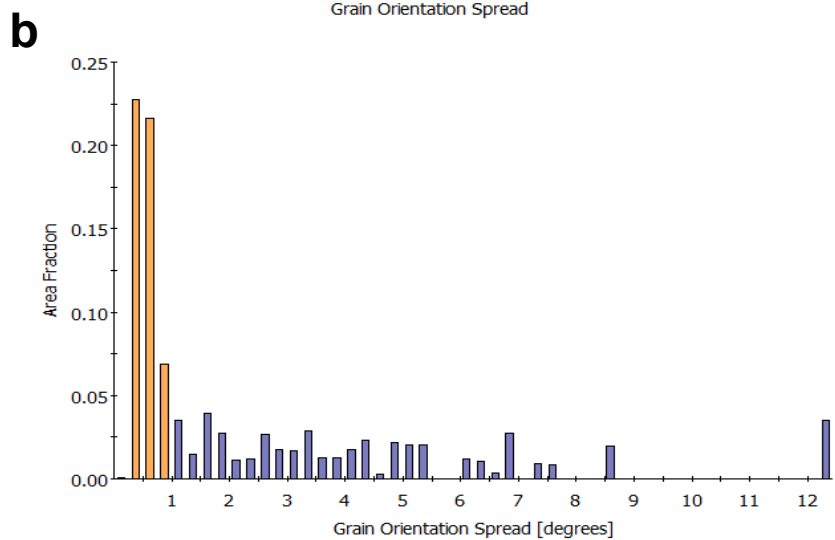

**Figure 4**: a) Representative EBSD phase maps for the UFH860-0.2s sample. Rx ferrite is shown in orange; non-Rx ferrite in blue; martensite in black and austenite is shown in white. HAGBs are represented in black and LAGBs in white. b) Histogram of grain orientation spread distribution in the ferrite matrix for the EBSD phase map presented in (a). Rx ferrite grains have GOS<1º (orange bars), whereas non-Rx ferrite grains have GOS>1º (blue bars), as described in Section 2.4.

**Table 1**: Data on the volume fraction of microstructural constituents as a function of heat treatment parameters.

| Peak temperature (ºC) | 860 | | 880 | | 900 | |
|---|---|---|---|---|---|---|
| Soaking time (s) | **0.2** | **1.5** | **0.2** | **1.5** | **0.2** | **1.5** |
| Martensite (%) | 6.9 ± 3.2 | 12.6 ± 3.1 | 11.6 ± 2.1 | 20.2 ± 2.5 | 16.3 ± 3.8 | 27.8 ± 4.6 |
| Retained austenite (%) | 2.2 ± 0.4 | 2.1 ± 0.3 | 1.2 ± 0.3 | 2.5 ± 0.7 | 1.7 ± 0.3 | 1.7 ± 1.0 |
| Ferrite (%) | 90.9 ± 4.0 | 85.3 ± 2.8 | 87.2 ± 1.9 | 77.3 ± 2.6 | 82.0 ± 3.6 | 70.5 ± 3.9 |
| *Rx ferrite* | 48.4 ± 9.8 | 61.8 ± 13.0 | 67.4 ± 3.8 | 73.5 ± 3.9 | 83.3 ± 3.7 | 80.3 ± 3.6 |
| *Non-Rx ferrite* | 51.6 ± 9.8 | 38.2 ± 13.0 | 32.6 ± 3.8 | 26.5 ± 3.9 | 16.7 ± 3.7 | 19.7 ± 3.6 |

Analysis of the effect of soaking time for each temperature shows that at 860 ºC, the martensite fraction increases from 6.9 % after 0.2 s to 12.6 % after 1.5 s. When temperature raises up to 880 ºC, the volume fraction of martensite formed after 0.2 s is 11.6 %, which increases to 20.2 % after 1.5 s. Finally, the 900 ºC treatment leads to the highest increment in martensite fraction during soaking, from 16.3 % after 0.2 s to 27.8 % after 1.5 s. For the shortest soaking time, the martensite volume fraction increases by the same amount (~4.7 %) when peak temperature is increased from 860 ºC to 880 ºC and then from 880 ºC to 900 ºC. Similar dependence can also be noted after soaking for 1.5 s. The ferrite volume fraction presents a reverse trend, as both phases are formed in the intercritical temperature range. The portion of RA is minor in all heat treated conditions, being between 1.2 to 2.5 %.

The morphology of the ferritic matrix is greatly affected by both parameters, temperature and soaking time. While at 860 ºC for 0.2 s the matrix is to larger extent formed by non-Rx ferrite (~ 52 %), whereas Rx ferrite prevails in the matrix after 1.5 s, reducing the volume fraction of the non-Rx grains to ~ 38 % (**Figure 5**). With increasing holding time at 880 ºC, the non-Rx fraction is reduced to a lesser extent from 33 % to 27 %. Finally, soaking at 900 ºC significantly reduces the volume fraction of non-Rx ferrite: Average volume fraction values are in the range of 17-20 % and are not affected by soaking time. Therefore, the most pronounced effect of soaking time in the studied temperature-time range occurs at the lower peak temperatures of 860 – 880 ºC, as seen from **Figure 5**.

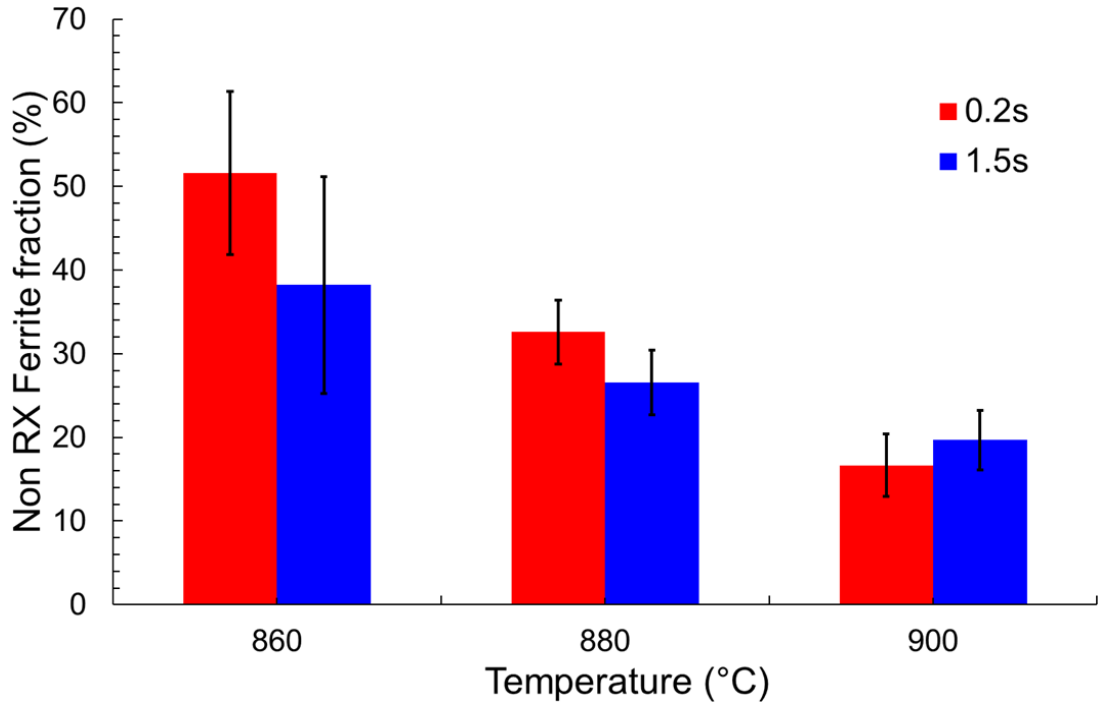

**Figure 5**: Volume fraction of non-Rx ferrite obtained from EBSD analysis for different temperatures (860, 880 and 900 ºC) and soaking times (0.2 and 1.5 s). In red for 0.2 s and in blue for 1.5 s.

The effect of the holding time (0.2 and 1.5 s) on the Rx and non-Rx ferrite grain size is shown in **Figure 6**. The effect of peak temperature after soaking for 0.2 s is shown in **Figure 6**a. First, the fraction of grains having size below 1 μm tends to decrease with increasing peak temperature. Second, after UFH to 860 ºC, the majority of grains have a size between 1 and 2 μm, although it tends to shift to higher values with peak temperature and reaching the range of 2 - 3 μm at 900 ºC. Third, there are some grains having a size above 6 μm even after heating to the lowest peak temperature of 860 ºC and their area fraction increases with peak temperature. The first two observations are even more pronounced when the holding time increases to 1.5 s (**Figure 6**b). It is shown that at 860 ºC, the fraction of grains below 1 μm has increased with respect to the 0.2 s counterpart, and it is considerably reduced at higher peak temperatures (880 ºC and 900 ºC). Moreover, the main fraction of grains presents a larger size when temperature is raised, although the fraction of grains larger than 6 μm has decreased for all temperatures at 1.5 s compared to the 0.2 s condition. On the other hand, the average grain size for the non-Rx ferrite at 0.2 s is higher compared to the Rx ferrite at all studied temperatures (**Figure 6** c), as the grains retained the initial cold rolled microstructure. In addition, the distribution is narrower compared to the Rx grains, as there are almost no grains below 1 μm. At the lowest temperature the distribution seems to be wider than at higher temperatures (880 and 900 ºC), although this is better seen after 1.5 s (**Figure 6** d). Coarse grains (> 6μm) are prone to disappear with temperature.

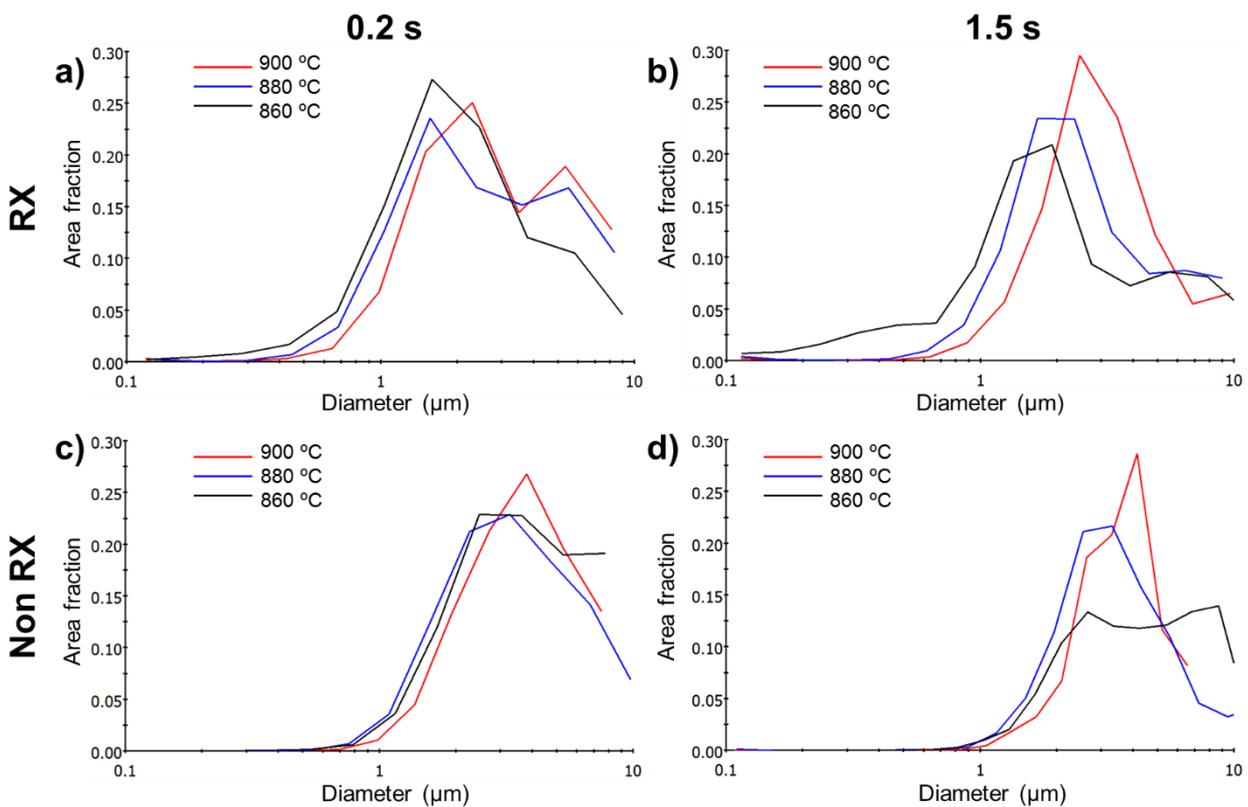

**Figure 6**: a) & b) representation of the area fraction for recrystallized (RX) ferrite grain size versus the equivalent circle diameter (ECD) vs. after 0.2 and 1.5 s holding time respectively for the different temperatures studied; c) & d) non-recrystallized (non-RX) ferrite grain size after 0.2 and 1.5 s holding time, respectively. Data are obtained from the EBSD measurements.

**Figure 7** represents the area fraction for martensite grains plotted versus the equivalent circle diameter (ECD) after soaking for 0.2 s and 1.5 s at the studied peak temperatures. For 860 and 880 ºC after 0.2 s (**Figure 7** a), the vast majority of martensite grains present an ultrafine size (below 1 μm). Contrary to the result seen in ferrite, the area fraction of grains < 0.5 μm increases when temperature is raised up to 880 ºC. However, when temperature further increases to 900 ºC, the curve is shifted to the larger grain size values, with the peak above 1 μm and showing a wider size distribution than at 860 or 880 ºC. Similar to ferrite, the effects are more pronounced with increasing soaking time to 1.5 s. At 860 ºC the austenite formed at intercritical temperatures grows, while at 880 ºC the fraction of grains below 1 μm increases resulting in a wider size distribution. Finally, at 900 ºC the increased intercritical austenite grain size shows a wider distribution than the 880 ºC and 860 º C.

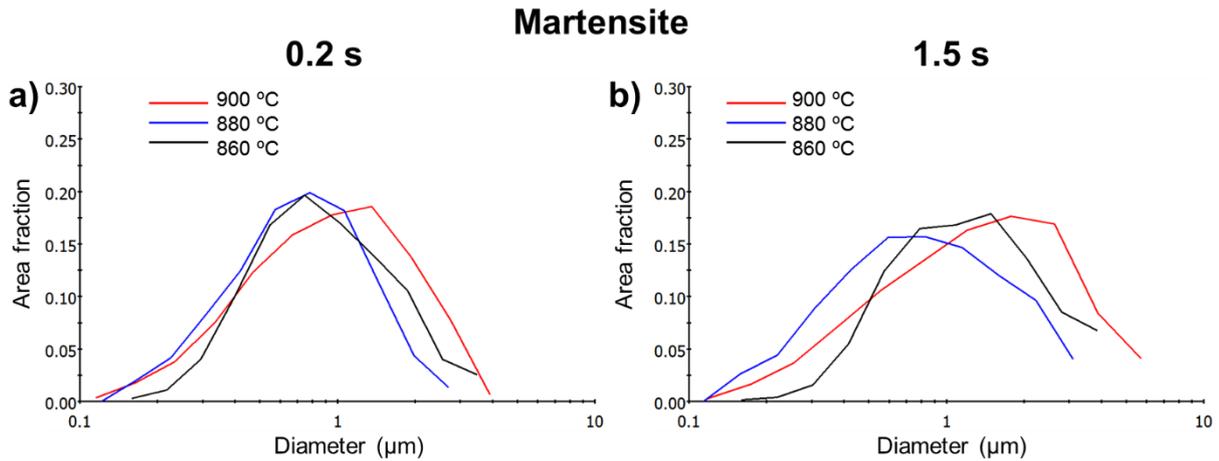

**Figure 7**: Martensite area fraction vs ECD after soaking for 0.2 s (a) and 1.5 s (b) at the peak temperatures of 860 ºC, 880 ºC and 900 ºC.

*3.2. Mechanical characterization*

*3.2.1. Properties of the individual microconstituents*

Nanoindentation tests were performed on selected grains to investigate the influence of maximum temperature and soaking time on the mechanical response of the microstructural constituents. **Figure 8**a shows an EBSD map where dark areas correspond to martensite and white areas to ferrite, which were later identified by SPM imaging prior nanoindentation testing (**Figure 8**b). An SPM image of the area with the nanoindentation imprints was also recorded immediately after the test as observed in **Figure 8**c. Finally, the microstructure was etched with nital 2 vol % to take SEM images (as shown in **Figure 8**d) that corroborate the differentiation made by EBSD. Typical load – depth curves for the main two microstructural constituents are shown in **Figure 8**e, where the red and blue curves correspond to martensite and Rx ferrite, respectively.

Continuous load-depth curves were obtained from nanoindentaion on martensitic grains, while majority of the ferritic grains exhibited pop-in events, particularly the softer Rx-ferrite grains. They are caused by sudden penetration bursts during the loading process. This effect has been related to the transition from an elastic to an elasto-plastic contact. The probability of pop-in events and the pop-in load increase as the dislocation density decreases, as discussed in our previous work [14].

The measured nanohardness values for the main microstructural constituents: Rx ferrite, non-Rx ferrite and martensite are summarized in **Table 2** as a function of peak temperature and soaking time. It is clearly seen that neither the soaking time nor the temperature affects the nanohardness of Rx ferrite, which has average values within the range of 2.5–2.6 GPa. Nevertheless, there is a significant difference between Rx and non-Rx ferrite, as the latter presents significantly higher nanohardness values (3.1–3.2 GPa), being also similar for all studied conditions. The martensite phase exhibits the highest nanohardness for all conditions, with the average values showing greater variation for each condition. For instance, after heating to 860 ºC and soaking for 0.2 s, the martensite average nanohardness value is 7.6 GPa, which reduces slightly to 7.4 GPa, when soaking increases to 1.5 s, but still within measured standard deviation. The softening effect with soaking time is much more evident at higher peak temperatures: at 880 ºC, the nanohardness drops by 10.3 %, varying from 6.8 GPa (after 0.2 s) to 6.1 GPa (after 1.5 s), while at 900 ºC, the drop is 16.7 %, as nanohardness varies from 6.6 GPa and 5.5 GPa for holding times of 0.2 and 1.5 s. Taking into consideration only the raise in temperature from 860 to 900 ºC, the average nanohardness drops by 13.1 % at 0.2 s (from 7.6 GPa to 6.6 GPa), while a more pronounced drop of 25.7 % was observed at 1.5 s (from 7.4 GPa to 5.5 GPa).

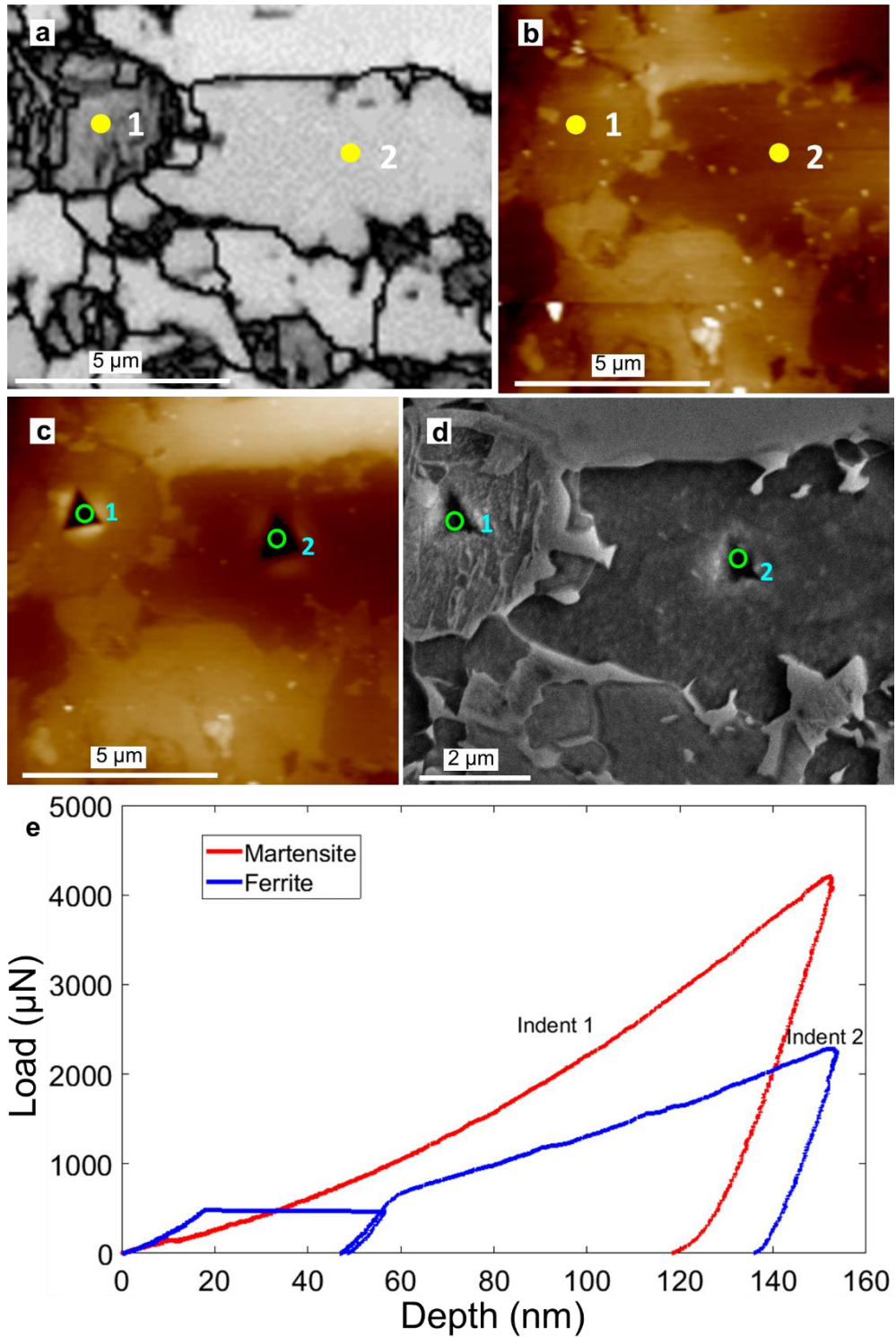

**Figure 8**: a) Band slope EBSD map with marked grains where the nanoindentation tests were performed; b) SPM map of the same area without indentation imprints; c) SPM map of the area after indentation; d) SEM image of the area after testing, etched with nital 2 vol.%; e) Typical indentation load – penetration depth curves from nanoindentation measurements on martensitic grain (in red color) and ferritic grain (in blue color).

Table 2: Data on nanohardness (in GPa) of the individual microconstituents.

| Peak temperature (ºC) | 860 | | 880 | | 900 | |
|---|---|---|---|---|---|---|
| Soaking time (s) | **0.2** | **1.5** | **0.2** | **1.5** | **0.2** | **1.5** |
| Martensite | 7.6 ± 2.4 | 7.4 ± 1.1 | 6.8 ± 0.8 | 6.1 ± 0.7 | 6.6 ± 1.0 | 5.5 ± 0.6 |
| Recrystallized ferrite | 2.6 ± 0.1 | 2.6 ± 0.1 | 2.6 ± 0.1 | 2.5 ± 0.1 | 2.6 ± 0.2 | 2.6 ± 0.1 |
| Non-recrystallized ferrite | 3.2 ± 0.2 | 3.1 ± 0.2 | 3.2 ± 0.2 | 3.2 ± 0.3 | 3.1 ± 0.2 | 3.2 ± 0.2 |

*3.2.2. Macro-mechanical characterization*

The effect of peak temperature and short soaking times on the macro-mechanical behavior of the AHSS was studied through hardness testing. The results are presented in **Table 3**. It is seen that an increment on the peak temperature produces a rise in the hardness values independently of the given soaking time. For 0.2 s, the increase in hardness between 860 ºC and 880 ºC is insignificant, going from 252 to 255 HV0.5 respectively, while at 900 ºC, the hardness increases to 264 HV0.5. The increase in hardness with peak temperature is much more evident for a soaking time of 1.5 s, being 245 HV0.5 at 860 ºC, and increasing to 272 and 293 HV0.5 at 880 and 900 ºC, respectively.

Table 3: Data on hardness of the heat treated strips.

| Peak temperature (°C) | 860 | | 880 | | 900 | |
|---|---|---|---|---|---|---|
| Soaking time (s) | 0.2 | 1.5 | 0.2 | 1.5 | 0.2 | 1.5 |
| Hardness (HV0.5) | 252 ± 4 | 245 ± 5 | 255 ± 5 | 272 ± 7 | 264 ± 5 | 293 ± 8 |

Additionally, tensile testing was carried out for all conditions using miniaturized dog bone samples. **Figure 9** illustrates the typical engineering stress - engineering strain curves. Data on the mechanical properties determined from the curves (0.2% proof strength $\sigma_{0.2}$, ultimate tensile strength $\sigma_{UTS}$, uniform elongation $\varepsilon_u$ and elongation to failure $\varepsilon_f$) are given in **Table 4**. One can see that at 860 °C, the yield strength slightly varies with soaking time being 444 MPa and 441 MPa for 0.2 and 1.5 s, respectively. However, for higher peak temperatures and holding times, the yield point is enhanced. For instance, at 880 °C, the $\sigma_{0.2}$ -values increase by more than 20 MPa to 468 and 473 MPa for 0.2 and 1.5 s, respectively. For the maximum peak temperature analyzed (900 °C), the yield strength for both soaking times shows the maximum values, being 479 MPa after 0.2 s and 493 MPa in the 1.5 s UFH treatment. Ultimate tensile strength values show a similar tendency than the yield point. While at 860 °C, the $\sigma_{UTS}$ is independent of the soaking time, with a value of ~925 MPa, higher peak temperatures and soaking times enhanced the strength of the material. For example, at 880 °C and after 0.2 s, the material presents a $\sigma_{UTS}$ of 933 MPa which increases up to 959 MPa at 900 °C. The increment in strength is more pronounced after longer soaking times, being 1017 and 1053 MPa for 880 °C and 900 °C respectively. Nevertheless, the uniform elongation shows the opposite trend, being reduced from 24 % to 15 % when the temperature is increased from 860 to 900 °C after 0.2 s. This reduction in elongation with temperature is less significant after 1.5 s, decreasing from 18 % at 860 °C to 15 % at 900 °C.

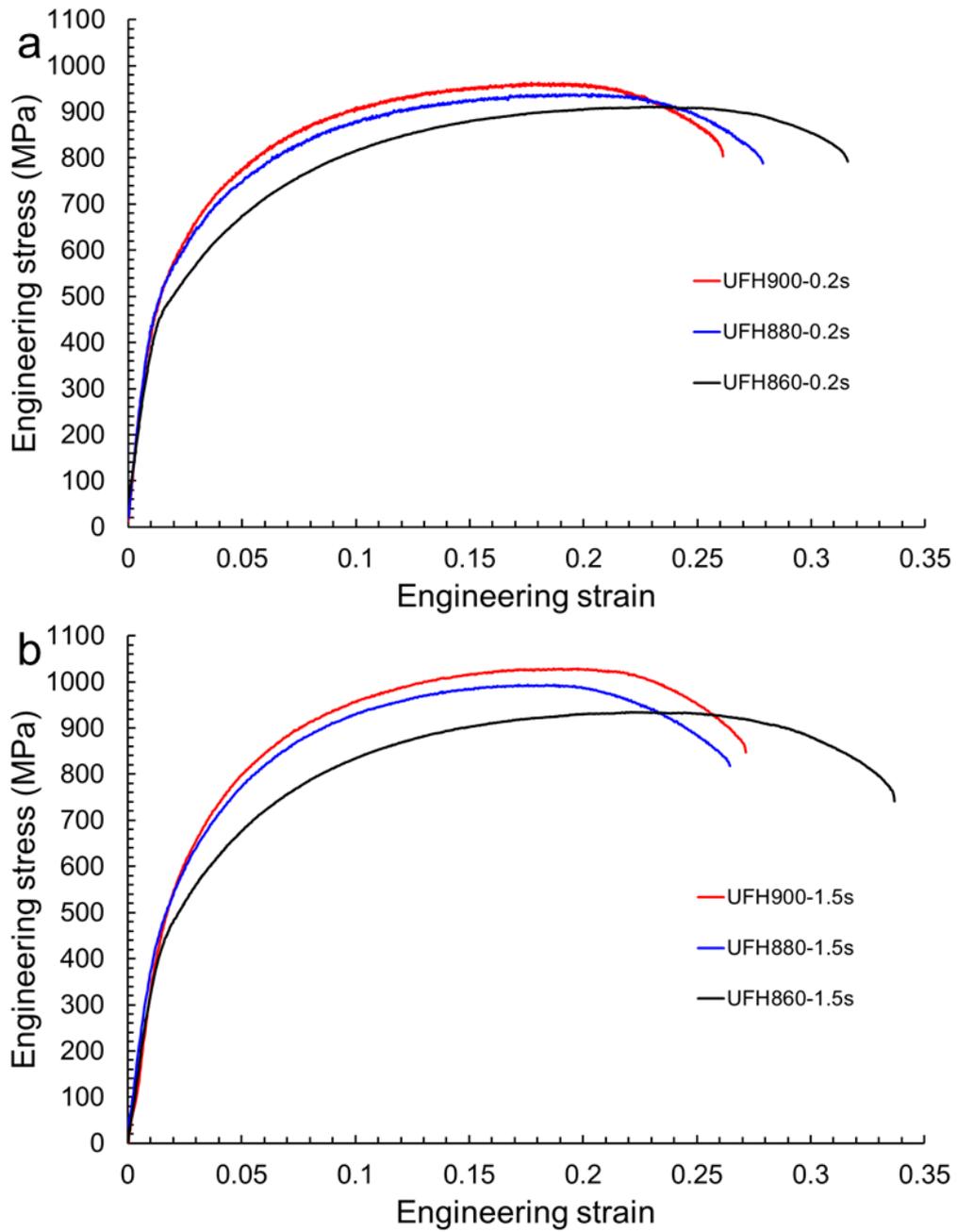

**Figure 9**: Typical engineering stress – engineering strain curves from tensile testing of specimens heated to different peak temperatures and soaked for: a) 0.2 s; b) 1.5 s.

**Table 4**: Basic mechanical properties determined by tensile testing of the heat treated samples.

| Peak temperature (ºC) | 860 | | 880 | | 900 | |
|---|---|---|---|---|---|---|
| Soaking time (s) | **0.2** | **1.5** | **0.2** | **1.5** | **0.2** | **1.5** |
| $\sigma_{0.2}$ | 444 ± 12 | 441 ± 8 | 468 ± 5 | 473 + 12 | 479 + 12 | 492 + 24 |
| $\sigma_{UTS}$ | 926 ± 20 | 925 ± 10 | 933 ± 5 | 1017 ± 28 | 959 ± 21 | 1053 ± 42 |
| $\varepsilon_u$ (%) | 24 ± 1 | 18 ± 2 | 18 ± 1 | 16 ± 2 | 15 ± 1 | 15 ± 1 |
| $\varepsilon_f$ (%) | 33 ± 1 | 30 ± 2 | 27 ± 1 | 24 ± 1 | 23 ± 1 | 23 ± 3 |

## 4. Discussion

*4.1. Influence of maximum temperature on the microstructure – properties relationship in the UFH steel*

Dilatometry tests (**Figure 2**) demonstrate that the higher the peak temperature, the higher the initial volume fraction of austenite, as it was reported elsewhere [29,30]. These results are in a good agreement with the outcomes of the EBSD analysis (**Table 1**), where the austenite/martensite fraction increases with temperature for the same soaking time. The variations between the initial austenite fraction measured for the studied temperatures through dilatometry and EBSD can be rationalized by the difference in the applied heating rate. While during dilatometry the maximum heating rate employed was 200 ºC/s, the samples analyzed by EBSD were processed at 800 ºC/s. It is known that high heating rates shift transformations temperatures ($A_{C1}$ and $A_{C3}$) to higher values [31,32]. Therefore, less amount of austenite is formed for the same peak temperature, when higher heating rates are applied. Moreover, heating at 800 ºC/s implies that the entire thermal treatment is faster than heating at 200 ºC/s, giving less time for the austenite nucleation process to be accomplished, resulting in the lower initial fraction of austenite observed in the samples heated at 800 ºC/s. Furthermore, both characterization techniques, dilatometry and EBSD, confirm that the increase in peak temperature favors the formation of austenite nuclei, as nucleation highly depends on temperature [15,33]. For instance, during the dilatometry test, intercricitical holding at 900 ºC results in a faster formation of austenite, and similar observations are made from the EBSD results analyzing the martensite grain size (**Figure 7**). In the latter case, the fraction of grains having size below 0.5 μm increases, when temperature is raised from 860 to 880 ºC for the shortest soaking time (**Figure 7**a). This means that the austenite nucleation is favored at 880 ºC, whereas at 860 ºC the already formed austenite tends to grow as the area fraction of grains above 1 μm is enlarged, being more evident after soaking for 1.5 s (**Figure 7**b). Hence, it is possible to state that at 860 ºC, the austenite growth is more significant than nucleation, whereas at 880 ºC, this behavior is inversed. At 900 ºC, both effects, nucleation and growth, are promoted, as the area fraction of grains below 0.5 μm and above 1 μm increase with respect to the 860 ºC case, due to the high internal energy. The rapid grain growth during ultrafast heating to high peak temperatures has been reported by Massardier *et al.* [8]. In addition to the grain size, the peak temperature also affects the interior structure of the formed austenite, which transforms into martensite after quenching. While at 860 ºC martensitic grains are chemically homogeneous, increasing the temperature and soaking time results in the appearance of non-homogeneous martensite regions, as it is shown in **Figure 3**e,f and **Figure 6**d which correspond to the UFH900-1.5s sample. Similar observations were reported by Castro *et al.* [17]. This effect can be rationalized based on the existence of carbon gradients in the grain interior, due to the high fraction of austenite formed, and the lack of time for carbon to diffuse through the grains [34].

Regarding the ferrite phase, it is possible to observe that UFH delays the recrystallization [7,11,14,15,17,35], leading to a matrix formed primarily by non-Rx ferrite, when the material is heated to 860 ºC for 0.2 s (**Figure 5**). The fraction of non-Rx ferrite significantly decreases with time and peak temperature, although it is not affected by the soaking time at the maximum peak temperature of 900 ºC and saturates at 17 – 20 %. The ferrite tends to transform into austenite at high temperatures, thus the driving force for recrystallization is reduced. Moreover, non-Rx grain size decreases with time and temperature (**Figure 6** c & d), favored by recrystallization and by the consumption of grains due to austenite formation [32]. Nevertheless, Rx ferritic grains show just the opposite behavior (**Figure 6**) growing with temperature and time. In addition, for low temperatures (860 ºC), ferrite tends to nucleate whereas, the temperature increment favors the ferrite grain growth, as the first nuclei formed rapidly grows [17]. This is related to the high stored energy from both, deformation induced via cold rolling and heat treatment.

*4.2.  Influence of the peak temperature on the mechanical behavior of the individual microconstituents*

Different microstructural constituents formed during heat treatment show dissimilar response during nanoindentation testing. For instance, Rx ferrite presents a lower nanohardness compared to the non-Rx ferrite independently of the heating rate and soaking time (**Table 2**). The latter exhibits large orientation gradients, as reported in our previous work [36], mainly associated to the high geometrically necessary dislocations (GND) density and residual stresses [37,38]. Moreover, it should be noted that the austenite to martensite transformation during quenching generates a volume expansion, which needs to be accommodated by the surrounding ferrite, introducing new dislocations in both, Rx ferrite and non-Rx ferrite [39]. The increment of the dislocation density can affect the ferrite mechanical behavior, as reported in [40]. The nanohardness of any of the ferrite microstructural constituents is not altered by the processing parameters (**Table 2**). On the other hand, the nanomechanical response of the martensitic grains is greatly affected by the processing parameters. At the lowest peak temperature (860 ºC) and shortest holding time (0.2 s), the martensite fraction is low (**Table 1**) due to the short time given to the austenite nuclei to form and grow. Thus, the carbon concentration within the austenite grains increases, due to the high fraction of ferrite [41]. As a consequence, the martensite grains formed at 860 ºC are the hardest (**Table 2**) in comparison to those formed at higher peak temperatures, as the hardness strongly depends on the carbon content [20,42]. Hence, the martensite strength is reduced with both, peak temperature and soaking time, due to the carbon homogenization in the austenite grains formed during the heat treatment. Similar results on the effect of holding time in DP steels and peak temperature during *Quenching & Partitioning* processing were reported by Mazaheri *et al.* [43] and Hidalgo *et al.* [44], respectively. The softening effect of increasing soaking times on the martensitic grains is more pronounced at higher peak temperatures due to the higher austenite fraction (**Figure 2**) and the more intensive grain growth (**Figure 7**). The latter also increases the diffusion distance of carbon [17] causing its redistribution inside the grains, reflected in the lower standard deviations of the measured nanohardness (**Table 2**).

*4.3. Relation of the peak temperature with the macro-mechanical behavior of the material*

The slight decrease in average hardness observed at 860 ºC when soaking time is increased from 0.2 s to 1.5 s (**Table 3**) is due to the higher fraction of Rx ferrite present at 1.5 s (**Table 1**), as Rx ferrite exhibits lower nanohardness compared to the non-Rx counterpart (**Table 2**). In addition, the presence of coarser grains at 1.5 s than at 0.2 s also leads to reduction in average hardness, obeying the Hall – Petch law [45]. However, raising the temperature to 880 or 900 ºC for 0.2 s leads to higher hardness due to the increased fraction of martensite [46]. On the other hand, holding times of 1.5 s at 880 and 900 ºC produce a notable increase in hardness compared to their 0.2 s counterparts, as a consequence of the considerable reduction of the ferrite volume fraction (**Table 1**). Similar results are observed for both, the yielding point and the ultimate tensile strength during tensile testing. At 860 ºC, when holding time is increased from 0.2 to 1.5 s, there is no variation of $\sigma_{0.2}$ or $\sigma_{UTS}$ (**Table 4**), although martensite fraction is increased with time (**Table 1**). This observation can be associated with the reduction of the non-recrystallized ferrite fraction, which presents a higher resistance to deformation compared to its recrystallized counterpart [47]. In addition, the grain size of ferrite also increases with soaking time resulting in a lower strength [45]. Nevertheless, the increment of temperature (880 ºC and 900 ºC) for a fixed soaking time strengthens the material (yield and ultimate tensile strength), due to a significant increase in the martensite volume fraction [48]. Our observations are consistent with the previous work published elsewhere [49,50]. Nevertheless, the increases in strength with both, temperature and soaking time, results in a significant loss in the ductility of the material. This is associated to the drop in the ferrite fraction, which is softer and more ductile than the martensite [51], which is also confirmed via nanoindentation (**Table 2**). In addition to the ductility, the strain hardening coefficient was analyzed for the different conditions, following the common power-law relationship in Eq. (1) [52]

$$\sigma = k\varepsilon^n \tag{1}$$

The strain hardening rate (n) in Eq. (1) was obtained using the following Eq.(2)

$$n = \frac{\ln\frac{\sigma_a}{\sigma_{a-1}}}{\ln\frac{\varepsilon_a}{\varepsilon_{a-1}}} \tag{2}$$

where $\sigma_a$ and $\varepsilon_a$ represent the true stress and true strain in the point *a,* respectively.

**Figure 10** shows the variation of the strain hardening rate with true strain for 0.2 and 1.5 s samples. For both holding times, at 860 ºC the material presents a higher strain hardening rate than its counterparts for any strain, decaying at a lower rate. This effect is associated with the higher fraction of non-Rx ferrite present in the microstructure, as compared to the ferrite formed at higher temperatures. In the non-Rx ferrite, the onset of plastic deformation requires higher stress (**Table 2**), due to its higher dislocation density. Therefore, when temperature or soaking time are increased, the strain hardening decreases, as a consequence of the reduction in the non-Rx ferrite fraction. When the microstructure shows a high martensite volume fraction, the difference in strain hardening rate is reduced. Several authors have associated this behavior to the martensite islands surrounded by ferrite. The volume expansion caused by the austenite to martensite transformation needs to be accommodated by the surrounding ferrite grains, resulting in strain hardening of the matrix [53,54].

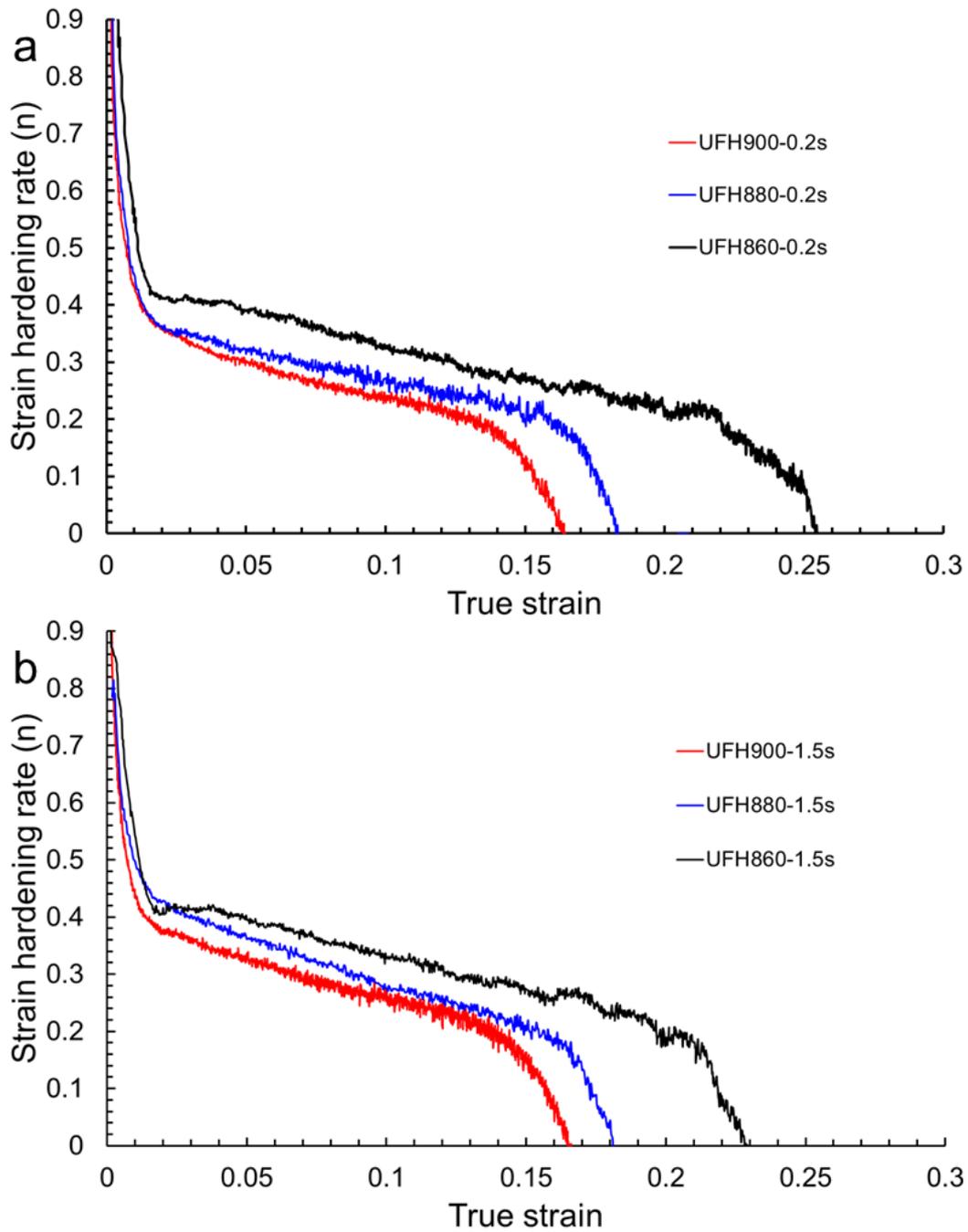

**Figure 10**: Representative strain hardening rate - true strain curves determined for samples treated at 860 ºC, 880 ºC and 900 ºC after soaking for 0.2 s (a) and 1.5 s (b).

*4.4. Potential microstructural and property gradients in the UFH processed sheets*

Analysis of the experimental results from the microstructural (Section 3.1) and mechanical (Section 3.2) characterization shows, that the industrial processing window of 20 ºC (i.e. ±10 ºC) should lead to some heterogeneity of the microstructure on the meso-scale (i.e. 0.1…1 mm). It will show up as some deviations in the size and local volume fraction of individual microstructural constituents (martensite, Rx ferrite and non-Rx ferrite), as well as the hardness of martensite. Nevertheless, such local heterogeneities of the microstructure should not degrade the overall mechanical behavior of the processed sheets on macro-scale. First, there are no significant differences in basic mechanical properties of the UFH processed steel within processing window of 20 ºC (**Table 4**). Second, the UFH processed steel shows high strain hardening ability independently of the peak temperature (see **Figure 9**, **Figure 10**). The latter should eliminate any minor negative effects from the microstructural heterogeneity in the UFH processed sheets appeared due to the deviation of the local peak temperature within ±10 ºC window.

## 5. Conclusions

We have studied the effect of the UFH parameters, peak temperature (860 ºC, 880 ºC and 900 ºC) and soaking times (0.2 s – 1.5), on the microstructure and mechanical response of a Fe-0.19C-1.61Mn-1.06Al-0.5Si steel at different scales. The main conclusions of our study include:

1) The increase in peak temperature promotes austenite formation. Independently of the peak temperature, mainly nucleation of small austenite grains with their limited growth occurs at the shortest soaking time (0.2 s), whereas both formation of nuclei and their growth proceed after soaking for longer time (1.5 s).

2) Morphology of the ferritic matrix is significantly altered by both peak temperature and soaking time. The lower peak temperature and shorter soaking time promote nucleation of the recrystallized ferritic grains, while the fraction of the non-recrystallized ferritic matrix undergoing recovery process remains high. With increasing both parameters, the average grain size of the recrystallized ferritic grains and their volume fraction tend to increase. These processes are accompanied by decrease of the ferrite volume fraction due to the increasing volume fraction of the intercritical austenite (i.e. martensite after quenching).

3) Independently on the applied heat treatment parameters, the highest nanohardness is measured on martensitic grains followed by non-recrystallized ferrite and recrystallized ferrite. Peak temperature and soaking time strongly affect the nanohardness of the martensitic grains. The higher the temperature the larger the grains, reducing the carbon concentration therein. On the contrary, nanohardness of the ferritic microconstituents is affected neither by temperature nor soaking time. The non-recrystallized ferrite is harder than its recrystallized counterpart due to the higher dislocation density of the former.

4) At the peak temperature of 860 ºC, the increase in soaking time within studied range does not produce an improvement in the mechanical properties despite higher martensite volume fraction, due to the decrease of the non-recrystallized ferrite fraction and the grain growth. Nevertheless, increasing peak temperatures to 880 ºC and 900 ºC favors the strengthening of the material, as the effect of martensite becomes the dominant factor. However, the ductility is considerably reduced with both, temperature and soaking time, due the lower fraction of the ductile ferritic phase.

5) Analysis of the experimental results from the microstructural and mechanical characterization shows that the industrial processing window of 20 ºC may lead to some heterogeneity in the microstructure of the UFH processed sheets. However, the latter should not have any negative effect on their overall mechanical behavior on the macro-scale.

**Acknowledgements**

MAT4.0-CM project funded by Madrid region under programme S2018/NMT-4381 is gratefully acknowledged by MAM, JMMA and IS. MAVT acknowledges gratefully the financial support by IMDEA Innovation Award. AK and RHP acknowledge funding from Materials Innovation Institute (M2i) from the project "UNDER" within the grant number s16042b.

**Data Availability**

The raw/processed data required to reproduce these findings cannot be shared at this time as the data also forms part of an ongoing study.